# Polarization Imaging and Edge Detection with Image-Processing Metasurfaces

MICHELE COTRUFO,[1,5,†] SAHITYA SINGH,[1,2,†] AKSHAJ ARORA,[1,2] ALEXANDER MAJEWSKI,[3] AND ANDREA ALÙ,[1,2,*]

[1]*Photonics Initiative, Advanced Science Research Center, City University of New York, New York, NY 10031, USA*
[2]*Physics Program, Graduate Center of the City University of New York, New York, NY 10016, USA*
[3]*Danbury Mission Technologies, LLC, an ARKA Group company, Danbury, CT 06810, USA*
[†]*The authors contributed equally to this work.*
[5]*mcotrufo@gc.cuny.edu*
[*]*aalu@gc.cuny.edu*

**Optical metasurfaces have been recently explored as ultrathin analog image differentiators. By tailoring the momentum transfer function, they can perform efficient Fourier-filtering – and thus potentially any linear mathematical operation – on an input image, replacing bulky 4f systems. While this approach has been investigated in different platforms, and several techniques have been explored to achieve the required angular response, little effort has been devoted so far to tailor and control also the polarization response of an image-processing metasurface. Here, we show that edge-detection metasurfaces can be designed with tailored polarization responses while simultaneously preserving an isotropic response. In particular, we demonstrate single-layer silicon metasurfaces yielding efficient Laplacian operation on a 2D image with either large polarization asymmetry, or nearly polarization-independent response. In the former case, we show that a strongly asymmetric polarization response can be used to unlock more sophisticated on-the-fly image processing functionalities, such as dynamically tunable direction-dependent edge detection. In parallel, metasurfaces with dual-polarized response are shown to enable efficient operation for unpolarized or arbitrarily polarized images, ensuring high efficiency. For both devices, we demonstrate edge detection within relatively large numerical apertures, with excellent isotropy and intensity throughput. Our study paves the way for the broad use of optical metasurfaces for sophisticated, massively parallel analog image processing with zero energy requirements.**

## 1. INTRODUCTION

Information and image processing is of paramount importance for several technologies and applications. Commonly, image processing is performed digitally, i.e., the image is digitized via a camera or a detector, and then processed with electronics and digital computation. While digital approaches are versatile and easy to implement, they also suffer from several drawbacks, such as high latency times, need of bias, and energy consumption, which are critical factors in several applications. These issues, combined with the exponentially growing demand for data processing [1], have renewed the interest in replacing digital data processing with analog optical computing [2]–[7] due to its appealing possibility of manipulating data at the speed of light while avoiding analog-to-digital conversion [8].

Analog image processing is conventionally performed via Fourier filtering techniques in the so called *4f* configuration, which requires two lenses with focal length f [9], [10]: the first lens performs an analog Fourier transform of the input image, creating a physical plane (located at a distance 2f from the input image) where each point corresponds to a different Fourier component. A spatially varying physical mask is then used to block certain Fourier components (depending on the target operation), while a second lens performs the inverse Fourier transform, rendering the output image at another 2f distance away. This *4f* approach, while easy to implement, is not suited for integrated devices because it is inherently bulky and prone to alignment issues. Notably, Fourier-based image processing can be implemented in a much more compact platform, by filtering the transverse momentum of an image directly in real space using metasurfaces [3], [6]. In particular, by engineering the nonlocal response of transversely invariant metasurfaces[11], [12], it is possible to perform momentum filtering within a very small footprint. Several studies [11]–[22] have demonstrated that optical metasurfaces can be used for this purpose, and the application that has garnered most attention is the use of these processing metasurfaces as compact analog differentiators – aimed at calculating the spatial gradients of an input image. The Laplacian operation — the second-order derivative of an input image $f_{out}(x,y) = (\partial_x^2 + \partial_y^2)f_{in}(x,y)$ — is indeed of particular interest due to its capability of enhancing the edges of an arbitrary input image (as sketched in Fig. 1a). In Fourier space, the Laplacian operation corresponds to a high-pass momentum filter, described by $f_{out}(k_x,k_y) = -(k_x^2 + k_y^2)f_{in}(k_x,k_y)$. To achieve this response, the metasurface must fully suppress plane waves incident at small angles, while progressively transmitting waves propagating at larger angles.

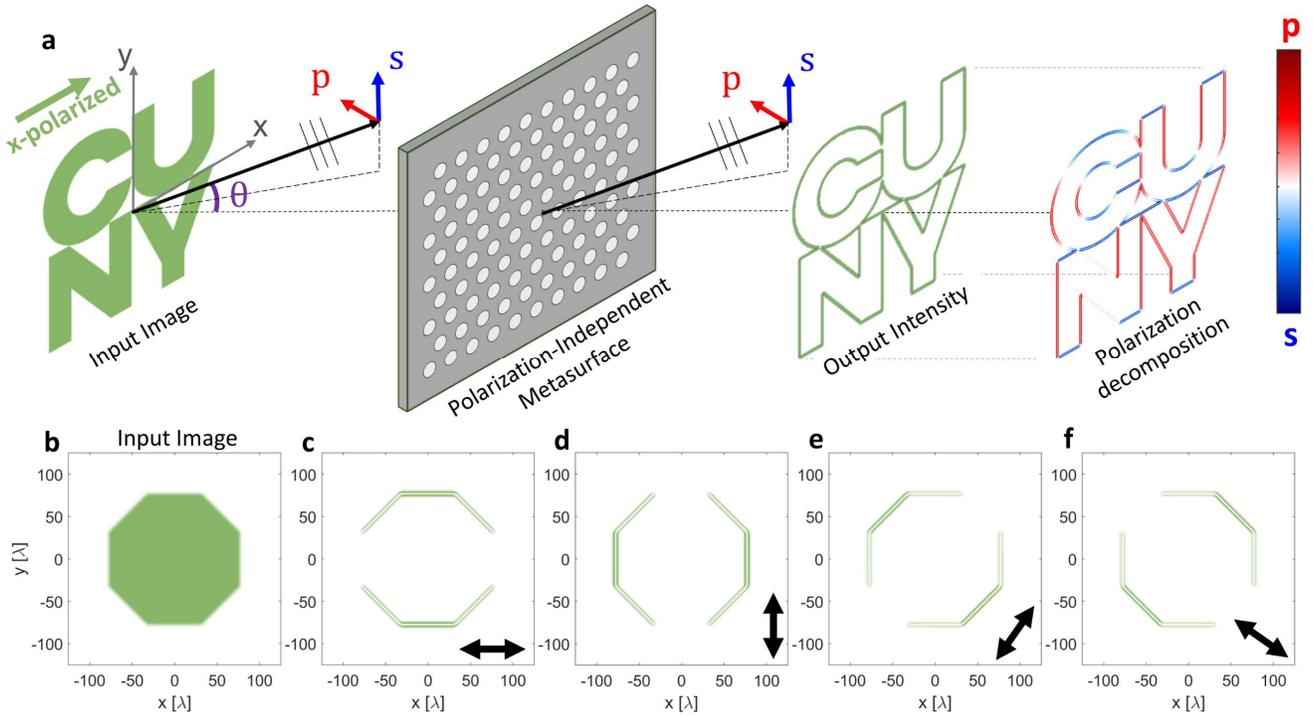

**Figure 1. (a)** Working principle of an edge-detection metasurface, showing the link between orientation of edges and the polarization of the transmitted image through the metasurface. In this panel we assume that an *x*-polarized input image (CUNY logo on the left) impinges on a polarization-independent edge-detection metasurface (gray device). The angular wave decomposition of the input image is composed of *s*- and *p*-polarized waves. While the output intensity field (green-colored plot labelled by 'Output Intensity') processed by the metasurface shows uniform and isotropic edge enhancement, each portion of the output image is in fact carried by waves with different polarization, as shown in the red-white-blue color-coded image in the right of panel a (see text for details). **(b-f)** Direction-dependent edge detection obtained via a polarization-dependent metasurface. An input image (b) impinges on a polarization-selective metasurface characterized by $t_{ss}(k_x,k_y) = (k_x^2 + k_y^2)$ and $t_{pp}(k_x,k_y) = t_{sp}(k_x,k_y) = t_{ps}(k_x,k_y) = 0$. In the output image (c-f), the edges are selectively enhanced depending on their orientation and the input polarization, denoted by the black arrow in each panel (c-f).

While this approach has been investigated and demonstrated in different platforms, and several methods have been explored to achieve the required angular responses, little effort has been devoted to engineering and controlling the polarization response of the metasurface, while simultaneously maintaining large efficiency, large numerical apertures and an overall isotropic response. Previous efforts [17], [18], [20], [21] have been focused on optimizing the response for some of these metrics, but neglecting the role that polarization can play to enhance the functionality. Here we show that light polarization control can be merged with momentum filtering over the same metasurface platform, and it can be exploited as an additional degree of freedom for image processing, leading to novel computational functionalities and to combining the concept of analog image processing with techniques such as polarization-difference imaging [23], [24] and polarimetry [25].

The presented metasurfaces enable full polarization control, while preserving high-efficiency and isotropic edge detection with tailored numerical aperture. In our work we leverage ultrathin silicon films, patterned periodically to engineer their nonlocal response, demonstrating that this platform can enable both strong polarization asymmetry and nearly polarization-independent responses. In the former case, this asymmetric response can be used to achieve on-demand direction-dependent edge detection tunable through the input polarization. Specifically, rather than enhancing all edges of an arbitrary image our metasurface enhances only the edges oriented along a desired direction, determined by the direction of the input polarization, and thus enabling polarization imaging functionalities dynamically controlled in real-time. Remarkably, this on-demand polarization response can be achieved within a single-layer metasurface design, compatible with CMOS fabrication, and, importantly, without the need of any additional polarizing element. At the same time, the device showcases a large numerical aperture (NA > 0.3), excellent isotropy, and large edge detection efficiency, close to the maximum limit for a passive device. Next, we show that the same metasurface platform can be optimized to obtain nearly polarization-independent edge-detection, which sacrifices the on-demand edge-detection control but enables larger efficiencies for unpolarized or arbitrarily polarized images. In this case, all edges are uniformly and isotropically enhanced for any combination of input polarizations.

## 2. RESULTS

**General Principle and Metasurface Design**

We begin by considering the general case of an input image angularly filtered by a metasurface (Fig. 1a), with the goal of elucidating how the output image depends on the interplay between the polarization of the input image, and the angular and polarization response of the metasurface. Assume an optical image in the plane z = 0 (Fig. 1a, left), described by an intensity profile $I_{in}(x,y) = |\mathbf{E}_{in}(x,y)|^2$, where $\mathbf{E}_{in}(x,y) = f_{in}(x,y)\mathbf{e}_{in}$ is the electric field with polarization direction $\mathbf{e}_{in} = \mathcal{E}_x \mathbf{e}_x + \mathcal{E}_y \mathbf{e}_y$ and angular frequency ω = 2πc/λ = $k_0$c. Following Fourier optics analysis [10], the image can be decomposed into a bundle of plane waves travelling in different directions, identified by the polar and azimuthal angles $(\theta, \phi)$, or equivalently by the in-plane wave vector $\mathbf{k}_\| = [k_x, k_y]$ with components $k_x = k_0 \sin\theta\cos\phi$ and $k_y = k_0 \sin\theta\sin\phi$. Each plane wave, propagating in a direction identified by $\mathbf{k}_\|$, can be decomposed into s- and p-polarized components according to [9]

$$\mathbf{E}_{in}(\mathbf{k}_\|) = \begin{pmatrix} E_s^{in}(\mathbf{k}_\|) \\ E_p^{in}(\mathbf{k}_\|) \end{pmatrix} = \tilde{f}_{in}(\mathbf{k}_\|) \begin{pmatrix} \frac{k_z}{k_0}(\hat{\mathbf{k}}_\| \times \mathbf{e}_{in})_z \\ \hat{\mathbf{k}}_\| \cdot \mathbf{e}_{in} \end{pmatrix}, \quad (1)$$

where $\tilde{f}_{in}(\mathbf{k}_\|)$ is the Fourier transform of $f_{in}(x,y)$, $k_z \equiv \sqrt{k_0^2 - |\mathbf{k}_\||^2}$, $\hat{\mathbf{k}}_\| \equiv \mathbf{k}_\|/|\mathbf{k}_\||$, $(...)_z$ denotes the z-component of the vector in the brackets, and an overall multiplicative factor has been omitted. After passing through the metasurface, the transmitted fields are

$$\mathbf{E}_{out}(\mathbf{k}_\|) = \tilde{f}_{in}(\mathbf{k}_\|) \begin{pmatrix} t_{ss}(\mathbf{k}_\|) & t_{sp}(\mathbf{k}_\|) \\ t_{ps}(\mathbf{k}_\|) & t_{pp}(\mathbf{k}_\|) \end{pmatrix} \begin{pmatrix} \frac{k_z}{k_0}(\hat{\mathbf{k}}_\| \times \mathbf{e}_{in})_z \\ \hat{\mathbf{k}}_\| \cdot \mathbf{e}_{in} \end{pmatrix}, \quad (2)$$

where $t_{ij}(\mathbf{k}_\|)$ are the complex co- and cross-polarized transmission coefficients. The fields in Eq. 2 represent the plane wave decomposition of the filtered output image, and the real-space output image $I_{out}(x,y) = |\mathbf{E}_{out}(x,y)|^2$ is obtained through the inverse Fourier transform (see [26] for additional details). Hence, the specific image processing performed by the metasurface depends on a nontrivial interplay between (i) the Fourier transform of the input image $\tilde{f}_{in}(\mathbf{k}_\|)$, (ii) the transfer functions of the metasurface $t_{ij}(k_x, k_y)$, and (iii) the input polarization $\mathbf{e}_{in}$. For example, in order to perform isotropic and polarization-independent edge detection, the transfer functions must satisfy $t_{ss}(k_x, k_y) = t_{pp}(k_x, k_y) \propto (k_x^2 + k_y^2)$ and $t_{sp}(k_x, k_y) = t_{ps}(k_x, k_y) = 0$. A schematic example of this behavior is shown in Fig. 1a. We assume that an x-polarized input image (CUNY logo, left part of the image) impinges on an isotropic and polarization-independent edge-detection metasurface (center), and we calculated [26] the intensity of the output image (green-coded color plot on the right of the metasurface). The output image intensity $I_{out}(x,y)$ displays clear, homogeneous and isotropic edge detection. Importantly, while the output intensity map shows uniform and isotropic edges, different components of the output image are actually carried (from the input image to the output image) by either s- or p-polarized waves. This is shown as blue-white-red color-coded lines in the right-most part of Fig. 1a (see [26] for details on how this plot is calculated). Here, each point of the output image is color-coded based on whether the waves that formed that specific point were mainly p-polarized (dark red) or s-polarized (dark blue). Clearly, the horizontal edges are exclusively carried by s-polarized waves, while the vertical edges are carried by p-polarized waves. Hence, by tailoring the polarization response of an edge-detecting metasurface, it is possible to selectively enhance only certain edges. Consider, for example, a metasurface with a strong polarization asymmetry, e.g., with all the transfer functions identically zero except for $t_{ss}(k_x, k_y) = (k_x^2 + k_y^2)$. The filtered electric field in Eq. 2 becomes

$$\mathbf{E}_{out}(\mathbf{k}_\|) = \tilde{f}_{in}(\mathbf{k}_\|) t_{ss}(\mathbf{k}_\|) \frac{k_z}{k_0}(\hat{\mathbf{k}}_\| \times \mathbf{e}_{in})_z \mathbf{e}_s. \quad (3)$$

Because of the term $\hat{\mathbf{k}}_\| \times \mathbf{e}_{in}$, all the Fourier components $\tilde{f}_{in}(\mathbf{k}_\|)$ corresponding to vectors $\mathbf{k}_\|$ parallel to the impinging polarization $\mathbf{e}_{in}$ will not be transmitted, thus not contributing to the output image. This effect becomes important when the input image $f_{in}(x,y)$ contains an edge oriented along a certain direction $\mathbf{n}$. In the limit of an infinitely long edge parallel to $\mathbf{n}$, [e.g., $f_{in}(x,y) = erf(\mathbf{n} \cdot \mathbf{r}/L)$, where $erf(x)$ is the error function and L is the length scale over which the edge intensity evolves from zero to one], the Fourier transform $\tilde{f}_{in}(\mathbf{k}_\|)$ is non-zero only when $\mathbf{k}_\|$ is orthogonal to $\mathbf{n}$. By combining this result with eq. 3, we find that

$$\mathbf{E}_{out}(\mathbf{k}_\|) = \tilde{f}_{in}(\mathbf{k}_\|) t_{ss}(\mathbf{k}_\|) \frac{k_z}{k_0}(\mathbf{n} \cdot \mathbf{e}_{in}) \mathbf{e}_s, \quad (4)$$

i.e., the edges parallel to the input polarization ($\mathbf{n} \cdot \mathbf{e}_{in} = 1$) contribute maximally to the filtered image, while the edges whose direction is orthogonal to the input polarization ($\mathbf{n} \cdot \mathbf{e}_{in} = 0$) will not contribute. The opposite scenario occurs if the metasurface works only for *p*-polarization, i.e., if only $t_{pp}(k_x, k_y) \neq 0$. In this case

$$\mathbf{E}_{out}(\mathbf{k}_\parallel) = \tilde{f}_{in}(\mathbf{k}_\parallel) t_{pp}(\mathbf{k}_\parallel) \; [(\mathbf{n} \times \mathbf{e}_{in})_z] \mathbf{e}_p. \qquad (5)$$

Now, edges whose direction is parallel to the input polarization ($\mathbf{n} \times \mathbf{e}_{in} = 0$) will not contribute to the output image. Equations 3-5 show that, by introducing polarization selectivity (i.e., by realizing a metasurface for which only $t_{ss}(k_x, k_y)$ or only $t_{pp}(k_x, k_y)$ provides the required Laplacian filtering, while the other is zero), it is possible to selectively enhance only the edges of an input image oriented along a desired direction, and that this direction is determined by the input polarization. Notice that this polarization selectivity can still be fully isotropic, which is important for imaging applications, enabling the detection of edges with a certain orientation independently of their position. To further numerically validate these results, we consider the case of an input image consisting of an octagon (Fig. 1b), impinging on a polarization-dependent edge-detection metasurface with only $t_{ss}(k_x, k_y) \neq 0$. Using the full expansion in Eqs. 1-2 (see [26] for more details), we numerically calculated [Figs. 1(c-f)] the intensity of the output image for different linear polarizations of the input image (denoted by the black arrow in each panel). The results match the expectations based on Eq. 4: for any input polarization, edges parallel to input polarization are maximally enhanced, while edges perpendicular to the input polarization are absent. We emphasize that the plots in Figs. 1(c-f) correspond to the *total* intensity of the output image. In other words, no additional polarization filtering is applied to the processed image, neither in these calculations nor later in our experiments. Instead, such strong polarization-dependent edge detection is embedded in the metasurface response.

## Experimental Results

We now demonstrate that edge-detection metasurfaces with either strongly asymmetric polarization response [e.g., $t_{ss}(k_x, k_y) \propto (k_x^2 + k_y^2)$ and $t_{pp}(k_x, k_y) = 0$] or a nearly polarization-independent response [i.e., $t_{ss}(k_x, k_y) \approx t_{pp}(k_x, k_y) \propto (k_x^2 + k_y^2)$] can be experimentally realized within a single-layer metasurface platform. We consider a thin silicon-on-glass metasurface operating in the near-infrared ($\lambda \approx 1500$ nm), but similar principles can be readily adapted to different materials and spectral ranges. The metasurface consists of a periodic triangular lattice of air holes (unit cell shown in Fig. 2a) etched into a silicon thin slab. As we show below, this same platform and unit cell design can be used to engineer vastly different polarization responses by simply varying three design parameters: the lattice constant *a*, the radius hole *R* and the slab thickness *H*. We begin by designing and demonstrating a metasurface with a strongly polarization-dependent response (Fig. 3). Using a parameter-sweep optimization, we identify a design ($a$ = 924 nm, $R$ = 265 nm, $H$ = 315 nm) that, up to an NA≈ 0.35, features the desired Laplacian-like response for *s*-polarization [$t_{ss}(k_x, k_y) \propto (k_x^2 + k_y^2)$], while it reflects any *p*-polarized wave [$t_{pp}(k_x, k_y) \approx 0$]. The metasurface was fabricated using standard lithographic techniques, as discussed in [26]. The normal incidence transmission spectrum of the fabricated device (Fig. 3a) features a broad transmission dip for wavelengths close to the operational wavelength $\lambda = 1490$ nm (shaded area in Fig. 3a). At this wavelength, the measured normal-incidence transmission is less than 1%, ideally suited to suppress the low-spatial-frequency components of the input image and hence enhance the edges. The metasurface response is polarization-independent at normal incidence, as expected from the $C_6$ symmetry of our design. However, the response at tilted angles is strongly asymmetric, as shown in Figs. 3(b-d). The measured *s*-polarized transmission amplitude (Fig. 3b, blue circles) features the desired quadratic increase versus $\sin\theta$, up to values of $\theta \approx 20^0$, as shown by the solid blue line in Fig. 3b which is a fit of the data with the function $t_{fit} = A(sin\theta)^2 + B$. Moreover, the transmission amplitude reaches values as high as $|t_{ss}| = 0.9$ (transmission ≈ 81 %), which is close to the upper bound set by the Fresnel reflection coefficient at the glass/air interface. In the same angular range, the *p*-polarized transmission amplitude remains very low, with a maximum transmission amplitude $|t_{pp}| \approx 0.12$ (transmission < 1.5%) at $\theta \approx 20^0$. This strong asymmetry is further demonstrated in Figs. 3(c-d), which show the measured amplitude of the transfer functions $|t_{ss}(k_x, k_y)|$ (Fig. 3c) and $|t_{pp}(k_x, k_y)|$ (Fig. 3d) for all impinging angles within a numerical aperture NA ≤ 0.5. Besides the strong polarization asymmetry, Fig. 3c confirms that the metasurface provides an excellent isotropic response up to NA = 0.34 (dashed circles in Figs. 3c and 3d), reproducing almost exactly the desired transfer function, $t_{ss}(k_x, k_y) \propto (k_x^2 + k_y^2)$. Next, we experimentally verify the edge-detecting behavior of this polarization-selective metasurface. In the experimental setup [22], [26], sketched in Fig. 2b, the input image is created by illuminating a target (consisting of the CUNY logo etched in a chromium mask) with a collimated quasi-monochromatic beam. The image created by the mask is collected by a NIR objective (Mitutoyo, 50X, NA = 0.42) placed on the opposite side, and relayed on a near-infrared camera with a tube lens. In this modality, the setup is essentially a standard optical microscope, and the camera records the unfiltered input image, shown in Fig. 3e. In order to perform analog edge detection, the metasurface is placed between the objective and the target (Fig. 2b). The filtered image recorded by the camera when the illumination is unpolarized (Fig. 3f) shows a clear, well-defined, isotropic and high-contrast edge enhancement. The quality of the detected edges can be further appreciated by the horizontal cut in Fig. 3g (corresponding to the dashed horizontal line in Fig. 3f), which shows that the intensity of the edges is almost 10x higher than the intensity of the background. Moreover, Figure 3e confirms that, even though the metasurface has a strongly asymmetric polarization response, under unpolarized excitation all edges are equally enhanced, independently of their orientation. This is expected from Eqs. 3-5: for unpolarized excitation the angular wave decomposition of each edge contains an equal mixture of s- and p-polarized waves, independently of the edge orientation. The p-polarized contribution is uniformly reflected by the metasurface, while the s-polarized contribution undergoes the desired Laplacian filtering, resulting in isotropic and direction-independent edge detection.

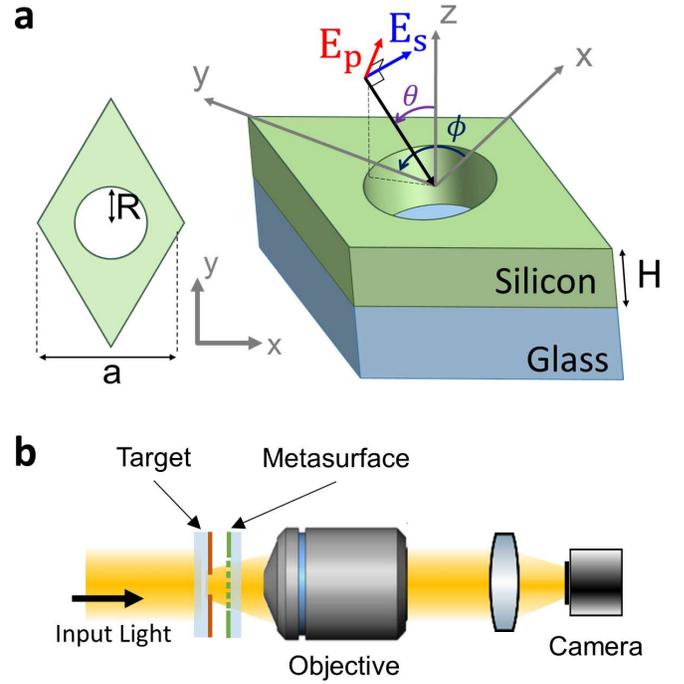

**Figure 2. (a)** Unit cell of the proposed metasurface. **(b)** Schematic of the experimental setup.

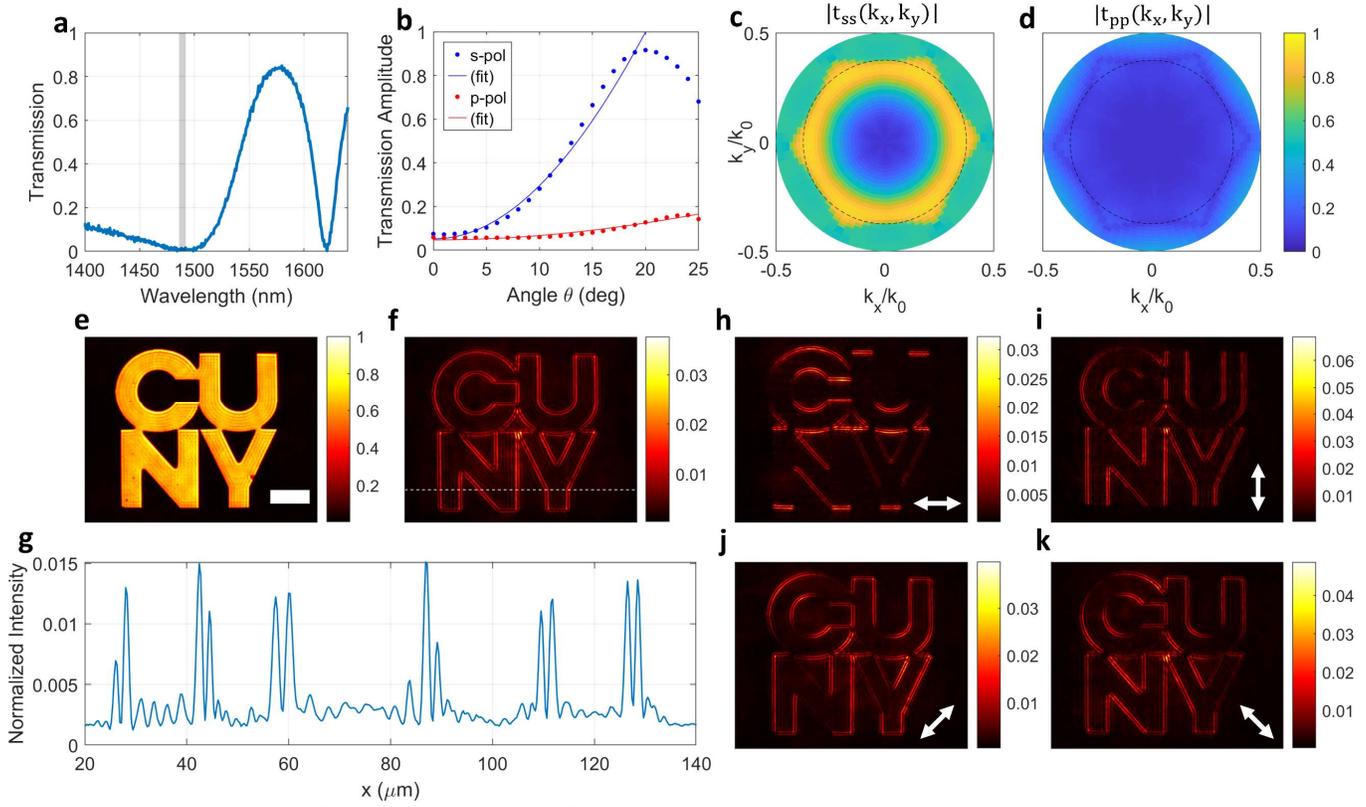

**Figure 3. Metasurface for polarization-dependent edge detection.** The unit cell of the metasurface is shown in Fig. 2a, with geometrical parameters $a$ = 924 nm, $R$ = 265 nm, $H$ = 315 nm. **(a)** Measured normal incidence transmission spectrum. The grey area denotes the operational frequency range. **(b)** Measured s- (blue) and p-polarized (red) co-polarized transmission amplitudes versus $\theta$, for fixed $\phi = 0$. Dots are experimental data, and solid lines are fits (see text for details). **(c-d)** Amplitude of the co-polarized $t_{ss}(k_x, k_y)$ (panel c) and $t_{pp}(k_x, k_y)$ (panel d) transfer functions of the metasurface. The dashed circles correspond to NA = 0.34 **(e)** Unfiltered test image, acquired with the setup in Fig. 2b without the metasurface (colorbar = 30 μm). **(f)** Output image when the metasurface is placed in front of the target, and the target is illuminated by unpolarized waves. **(g)** Horizontal cut corresponding to the white dashed line in (f). **(h-j)** Same as (f), but with polarized excitation: (h) horizontal, (i) vertical, (j) diagonal, (k) anti-diagonal polarization.

The measured response drastically changes when the impinging illumination is, instead, polarized. As shown in Figs. 3(h-k) for different linear polarizations, only the edges whose direction $\mathbf{n}$ has a nonzero component parallel to the impinging polarization $\mathbf{e}_{in}$ are enhanced, following the intensity trend $I \propto |\mathbf{n} \cdot \mathbf{e}_{in}|^2$ dictated by Eq. 4. In particular, only horizontal (vertical) edges are enhanced when the illumination is horizontally (vertically) polarized, as shown in Figs. 3h and 3i. As discussed above, this trend follows from the fact that the designed metasurface has a high efficiency Laplacian-like response for s-polarization [ $t_{ss}(k_x, k_y) \propto (k_x^2 + k_y^2)$ ], while it completely suppresses p-polarization [ $t_{pp}(k_x, k_y) = 0$ ]. If we designed a metasurface with opposite response (i.e., where the role of s- and p-polarization are reversed), the opposite behavior would be observed, i.e., only horizontal edges would be enhanced under vertically-polarized illumination. The strong contrast between the intensity of edges oriented along orthogonal directions can be further verified by repeating the experiment with rectangular shapes as input images, as shown in [26]. As recently pointed out [22], an important figure of merit for edge-detection metasurfaces is their intensity efficiency, that is, how the intensity of the output image compares to the intensity of the input image. Following [22], we consider the peak efficiency defined as $\eta_{\text{peak}} \equiv \max(I_{\text{out}})/\max(I_{\text{in}})$, where $\max(I_{\text{in/out}})$ are the maximum intensities in the input and output images. In order to readily quantify this efficiency, all the experimental images in Figs. 3(e-k) have been normalized by dividing the counts recorded in each camera pixel by the camera integration time and the power impinging on the sample. This allows us to correctly compare the energy flux impinging on the camera in different scenarios. Moreover, for numerical convenience the intensities of all images have been further renormalized such that the maximum intensity of the input image (Fig. 3e) is 1. As a result of this normalization procedure, the peak efficiency $\eta_{\text{peak}}$ can be readily extracted from the maximum values of the colorbars of the filtered images, leading to typical values of $\eta_{\text{peak}} \approx 3\% - 7\%$. These values are in line with values measured in [22] and, importantly, they are quite close to the maximum efficiency obtainable for any ideal passive edge-detection device [22]. These significantly large efficiencies are a direct consequence of the large transmission ($|t_{ss}|^2 \sim 81\%$) enabled by our metasurface at large angles (see Figs. 3b-c), which is achieved thanks to the optimized design, the simple fabrication process, and the absence of absorbing materials.

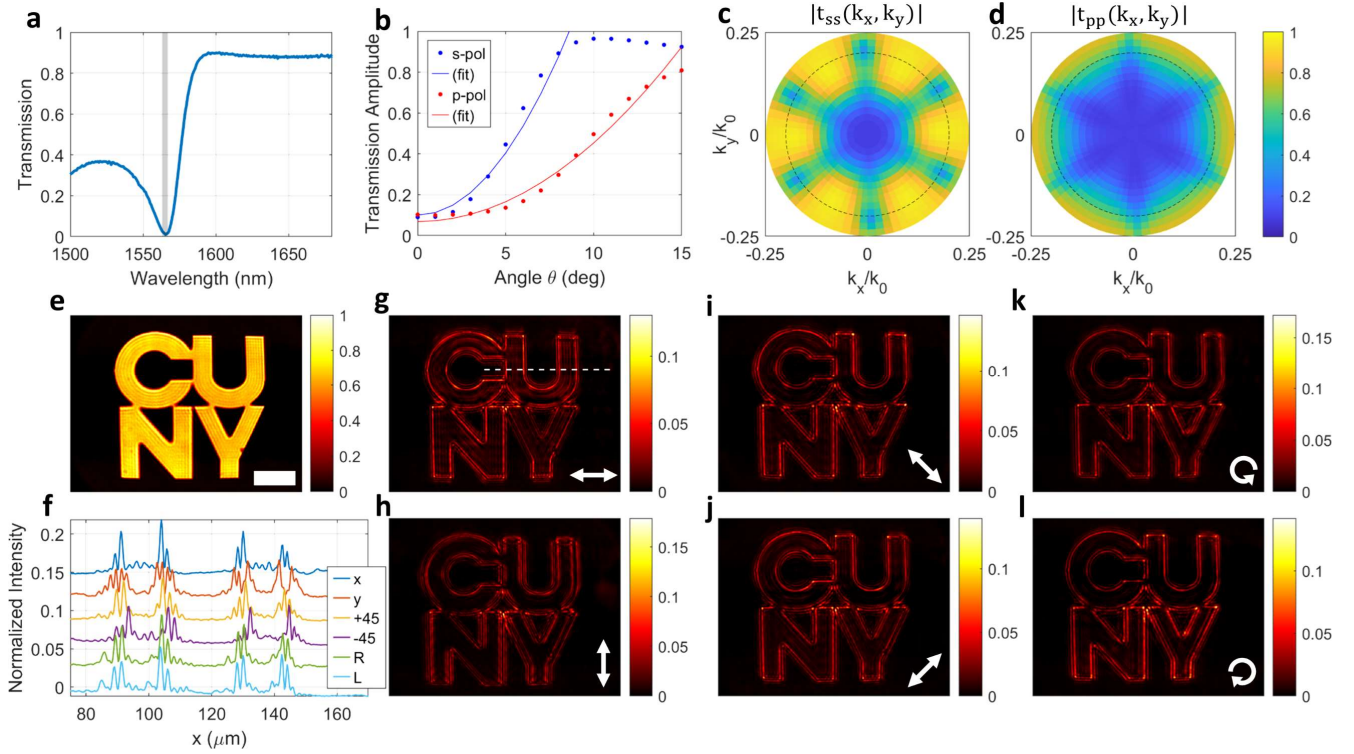

**Figure 4. Metasurface for polarization-independent edge detection.** The unit cell of the metasurface is shown in Fig. 2a, with geometrical parameters $a$ = 785 nm, $R$ = 153 nm, $H$= 273 nm. **(a)** Measured normal incidence transmission spectrum. The grey area denotes the operational frequency. **(b)** Measured s- (blue) and p-polarized (red) co-polarized transmission amplitudes versus $\theta$, and for fixed $\phi = 0$. Dots are experimental data, and solid lines are fits (see text for details). **(c-d)** Amplitude of the co-polarized $t_{ss}(k_x, k_y)$ (panel c) and $t_{pp}(k_x, k_y)$ (panel d) transfer functions of the metasurface. **(e)** Unfiltered test image, colorbar = 30um. **(f-l)** Experimentally measured edge detection. Panels (g-l) show the filtered image when the metasurface is placed in front of the target, and the target is illuminated by light with different polarization: horizontal (g), vertical (h), anti-diagonal (i), diagonal (j), right circular (k) and left circular (l). Panel (f) shows horizontal cuts from (g-l), corresponding to the segment denoted by the dashed horizontal line in panel g. Plots are displaced vertically in steps of 0.03.

As demonstrated above, an edge-detection metasurface with a strongly asymmetric polarization response can be used to achieve direction- and polarization-dependent edge detection. For other applications, however, a polarization-independent response [ $t_{ss}(k_x, k_y) \approx t_{pp}(k_x, k_y) \propto (k_x^2 + k_y^2)$ ] may be desirable. Indeed, a polarization-independent response guarantees isotropic and homogeneous edge detection for any impinging polarization, including unpolarized light. Moreover, a polarization-independent response is also beneficial to increase the output intensity (and thus the efficiency $\eta_{\text{peak}}$) when the input image is of arbitrary polarization. For instance, under unpolarized excitation the polarization-dependent metasurface in Fig. 3 automatically rejects approximately half of the input intensity (corresponding to p-polarized waves), thus strongly reducing the overall efficiency. Remarkably, we show here that the same metasurface platform used for the previous design – a triangular lattice of air holes etched in a silicon thin film – can also be used to obtain a polarization-independent response. To achieve this, we performed a second optimization by varying the three parameter designs ($a$, $R$ and $H$). In the optimization, we looked for designs for which, at a given wavelength, the transmission is zero at normal incidence and, simultaneously, the $s$- and $p$-polarized transmissions are large and similar at larger angles. The optimized design ($a$ = 785 nm, $R$ = 153 nm, $H$ = 273 nm) was fabricated and tested with the same experimental procedures used for the single-polarization device shown in Fig. 3. The normal-incidence spectrum of this device (Fig. 4a) features a narrower transmission dip at $\lambda \approx 1560$ nm, which defines the operational wavelength. The angle-dependent transmission measurements (Figs. 4(b-d)) confirm that the $s$- and $p$-polarized transfer functions feature the required quadratic-like increase with $sin\theta$, albeit within a narrower numerical aperture and with lower isotropy compared to the response obtained in Fig. 3c. Despite these potentially detrimental issues, the imaging experiments [Figs. 4e and 4(g-l)], performed in the same conditions and with the same target as in Fig. 3, confirm that this device leads to high-quality, highly isotropic, and polarization-independent edge detection. For any linear (Figs. 4(g-j)) and circular polarization (Figs. 4(k-l)) input, all edges of the input image are equally enhanced, and independently of their orientation. This is further confirmed by the plots in Fig. 4f, corresponding to horizontal cuts of Figs. 4(g-l) at the position marked by the dashed white line in Fig. 4g. Moreover, as expected from the discussion above, this dual-polarization device displays an increased efficiency with respect to the single-polarization device shown in Fig. 4. The peak efficiency $\eta_{peak}$, extracted from the upper limits of the colorbars in Figs. 4(g-l), is above 10% for all input polarizations.

## 3. DISCUSSION AND CONCLUSIONS

We have demonstrated full-control over the polarization response of edge-detection metasurfaces, serving as a new knob to tailor analog image processing on demand using metasurfaces. In particular, our work shows that, by using conventional materials and design platforms – in our case a single-layer patterned thin silicon film – it is possible to realize edge-detection metasurfaces that exhibit either a strongly asymmetric polarization response, or a nearly polarization-independent response, while maintaining a high degree of isotropy. In the former case, the device imparts the required Laplacian-like response only for s-polarized impinging waves, while it completely suppresses p-polarized waves. We demonstrated that this functionality can be used to achieve controllable direction-dependent edge detection: only the edges parallel to a certain direction, determined by the input polarization, are enhanced. This functionality paves the way to applications in the fields of polarization-difference imaging [23], [24] and polarimetry [25]. We then showed that, by simply tweaking a few geometrical parameters, a nearly polarization-independent device can be obtained. This nontrivial feature allows performing high-contrast edge detection independently of the input polarization and edge orientation, increasing the edge-detection efficiency for unpolarized images.

More generally, our results demonstrate that, by leveraging the interplay between polarization of the input image and polarization response of the metasurface, it is possible to unlock novel functionalities in the field of analog image processing. We expect our work to pave the way towards a novel class of devices where nontrivial computational tasks can be achieved by further engineering the interplay between input polarization and the metasurface angular- and polarization-response.

**Funding.** This work was supported by Danbury Technologies, the Air Force Office of Scientific Research MURI program and the Simons Foundation. Device fabrication was performed at the Nanofabrication Facility at the Advanced Science Research Center at The Graduate Center of the City University of New York

**Competing interests.** The authors declare no conflicts of interest.

**Contributions.** All authors conceived the idea and the corresponding experiment. A.Ar. and S.S. performed numerical simulations and optimizations. M.C. fabricated the devices and performed the experimental measurements together with S.S.. A.Al. supervised the project. All authors analyzed the data and contributed to writing the manuscript.

**Corresponding author.**
Correspondence to Andrea Alù.

**Data availability.** Data underlying the results presented in this paper may be obtained from the authors upon reasonable request.

# Polarization Imaging and Edge Detection with Image-Processing Metasurfaces: supplemental document

Michele Cotrufo,[1,5,†] Sahitya Singh,[1,2,†] Akshaj Arora,[1,2] Alexander Majewski,[3] and Andrea Alù,[1,2,*]

### S.1 Sample Fabrication

The samples were fabricated by following the same approach as in ref. [1]. Commercially available glass coverslips (25 x 75 x 1 mm, Fisher Scientific) were used as transparent substrates. The substrates were cleaned via an acetone bath inside an ultrasonic cleaner followed by an oxygen-based cleaning plasma (PVA Tepla IoN 40). After cleaning, amorphous silicon (α-Si) was deposited on the substrates via a plasma-enhanced chemical vapor deposition (PECVD) process. A layer of E-beam resist (ZEP 520-A) was then spin-coated on top of the samples. To avoid any charging effect during electron-beam lithography, a thin layer of an anti-charging polymer (DisCharge, DisChem) was additionally spun on the sample. The desired pattern was then written with an electron beam tool (Elionix 50 keV). The anti-charging layer was removed with water, and the exposed ZEP was developed with n-amyl acetate followed by a rinse in isopropanol. The pattern was then transferred to the underlying silicon layer via dry etching in an ICP machine (Oxford PlasmaPro System 100). The resist mask was finally removed with a solvent (Remover PG).

### S.2 Optical Characterization

The transmission measurements shown in Figs. 3(a-d) and Figs. 4(a-d) of the main text were performed with the custom-built setup shown in Fig. S1a. To measure the transmission spectra of the devices (Figs. 3a and 4a), a fiber-coupled broadband light (Thorlabs, SLS201L) was linearly polarized and weakly focused on the sample via a lens (L1) with f = 20 cm focal length. This resulted in an excitation spot with a diameter of about 500 $\mu$m. The transmitted signal was collected and re-collimated on the other side of the sample by an identical lens (L2), and then focused on the aperture slit of a near-infrared spectrometer (OceanOptics, NIRQuest), denoted by P2 in Fig. S1a. We used this approach to measure the normal-incidence transmission spectra, *i.e.,* with $\theta = 0°$ in Fig. S1a.

To measure the angle-dependent transfer functions [Figs. 3(b-d) and Figs. 4(b-d)], the same setup was used but with different sources and detectors. Additionally, the sample was mounted on two different rotation stages, a motorized one (Thorlabs, HDR50) to control the polar angle $\theta$, and a manual one to control the azimuthal angle $\phi$. A broadband supercontinuum laser (NKT, SuperK) was filtered either via a commercial narrowband filter (Photon, LLTF Contrast) or with a custom-based pulse shaper, and then injected into the setup via a fiber. When using the commercial filter, the excitation has a linewidth of approximately 5 nm (see Fig. S1c). When using the custom-made pulse shaper as a filter, the excitation linewidth can be reduced to approximately 3 nm (see Fig. S1d). A beam splitter (BS in Fig. S1a) was used to redirect approximately 50% of the laser power to a germanium photodiode P1 (Thorlabs, S122C). A linear polarizer placed before the beamsplitter was used to polarize the incoming beam along either x or y, which correspond, respectively, to p- and s-polarization for any value of $\theta$ and $\phi$. The laser was focused on the sample and collected on the other side using the same lenses L1 and L2 used for the spectrometer-based measurements of the transmission spectra. The transmitted signal was measured by another identical germanium powermeter (P2). The angle-dependent transmission was then measured obtained by sweeping the angles $\theta$ and $\phi$, and recording the corresponding powers measured by P1 and P2. An additional calibration run was taken without the metasurface, to account for the exact splitting ratio of the BS and for discrepancies between the two powermeters.

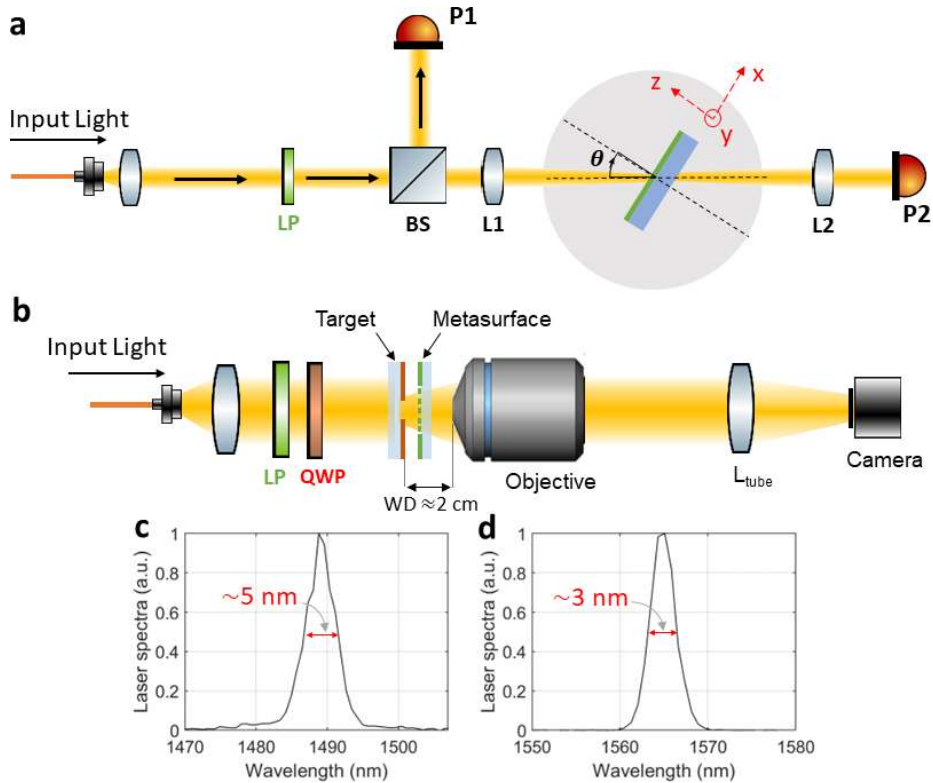

**Figure S1. (a)** Schematic of the setup used to perform the angle-dependent transmission measurements. LP = Linear polarizer, BS = beamsplitter, L1, L2 = lenses, P1, P2 = photodiodes. See text for additional details. **(b)** Schematic of the setup used for the imaging experiments. **(c-d)** Representative spectra of the excitation used for the experiment with the polarization-dependent device (panel c) and with the polarization-independent device (panel d).

The imaging experiments shown in Figs. 3 and 4 of the main paper were performed with the setup shown in Fig. S1b. The illumination was provided by the same filtered supercontinuum source used in the angle-dependent transmission measurements. The excitation was collimated and prepared in the desired polarization state by a linear polarizer (LP) and quarter wave plate (QWP). For the measurements with unpolarized excitation (Fig. 3f), all polarization optics (LP and QWP) were removed. The output of the commercial filter provides a nearly-unpolarized source, as confirmed by the measured degree of polarization < 7%. The collimated beam impinges on aa target consisting of a desired shape (e.g. the CUNY logo) etched into a chromium layer, which creates the input image. The image created by this mask is collected by a NIR objective (Mitutoyo, 50X, NA = 0.42, working distance = 2 cm) placed on the opposite side, and relayed on a near-infrared camera (MKS/Ophir, SP1203) with a tube lens (f = 15 cm).

In this modality, the setup is essentially a standard optical microscope, and the camera records the unfiltered input image (i.e. Figs. 3e and 4e). In order to perform analog edge detection, the metasurface is placed between the objective and the target.

**S.3 Edge detection with rectangular targets**

In Fig. 3 of the main text we demonstrated direction-dependent edge detection using a CUNY logo as a test image. In Fig. S2 we show additional imaging measurements done with rectangular shapes. All experimental conditions, including the excitation spectrum, are the same as in Fig. 3 of the main text. Fig. S2a shows the unfiltered image, containing three rectangular shapes with width of 100 μm and different heights. Figs. S2(d-g) show the filtered image (with the same metasurface used in Fig. 3 of the main paper) as a function of the input polarization, as indicated in the right-bottom corner of each image. The results clearly confirm that for x-polarized excitation (Fig. S2d) only the horizontal edges of the rectangle are visible, while the vertical ones are completely absent. This is further shown in Figs. S2(b-c), where we plots the intensity profiles along the horizontal dotted-dashed line and the vertical dashed line in Fig. S2d. For y-polarized excitation (Fig. S2e), instead, only vertical edges are visible while the horizontal ones are absent. For polarizations oriented along either the diagonal (Fig. S2f) or anti-diagonal (Fig. S2g) all edges are equally enhanced, since in this case the input polarization contains an equal mixture of x- and y-polarized electric field.

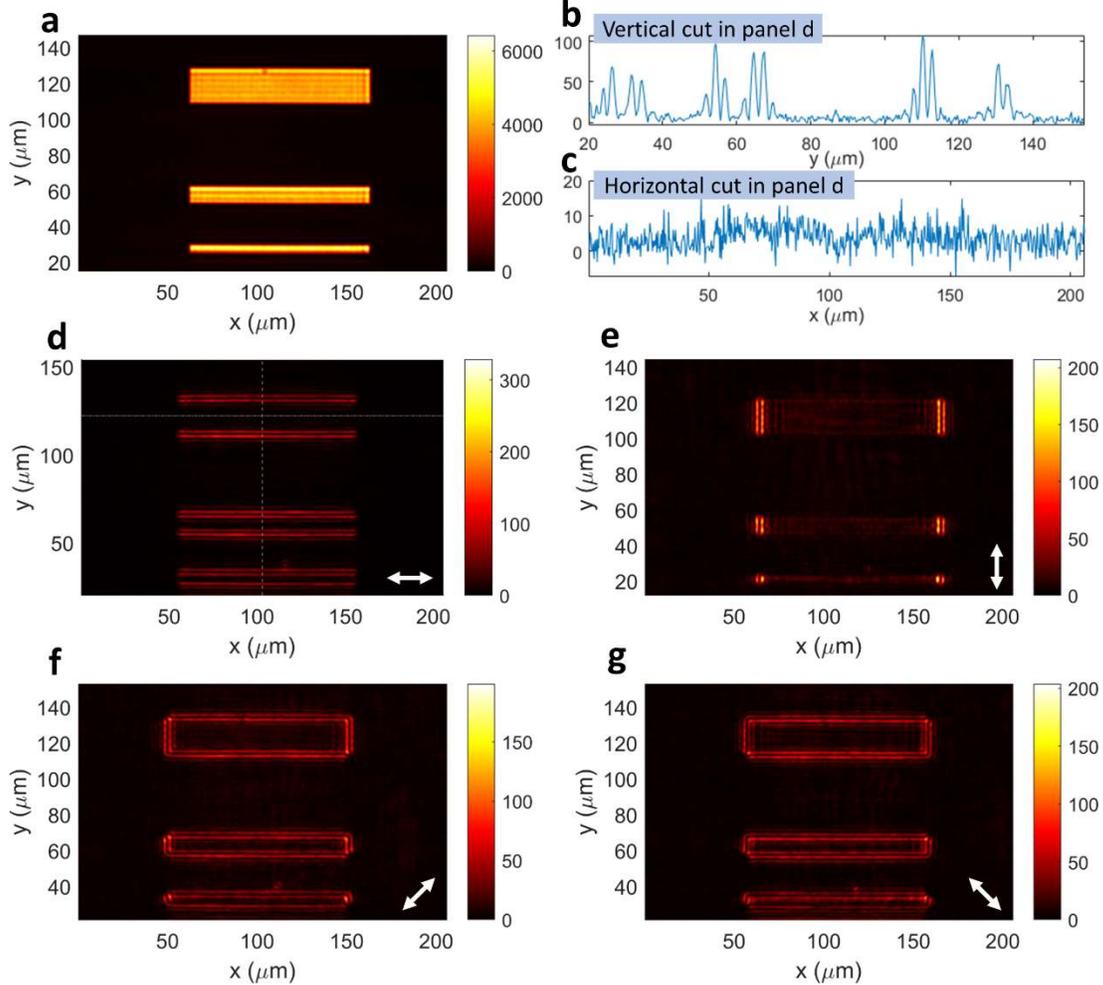

**Figure S2. Direction-dependent edge detection with rectangular targets. (a)** Unfiltered image. **(b-c)** Intensity profiles along the vertical (panel b) and horizontal (panel c) lines in panel d. **(d-g)** Output images when the metasurface is placed in front of the target and for four different linear polarizations of the input light: (d) x, (e) y, (f) the x-y diagonal, (g) the x-y anti-diagonal.

### S.4 Image processing with metasurfaces – theoretical and numerical calculations

In this paragraph we outline the details of the mathematical steps required to calculate the image processing imparted by a generic metasurface, and describe the procedure used to generate the different panels of Fig. 1 of the main text. The derivation follows closely the one presented in the supplementary materials of ref. [1]. We begin by assuming (Fig. S3) that an optical image is defined in the plane z = 0 by a spatially varying intensity profile $I_{in}(x,y) = |\mathbf{E}_{in}(x,y)|^2$, where $\mathbf{E}_{in}(x,y) = E_{in}(x,y)\mathbf{e}_{in}$ is an electric field with electric field polarization $\mathbf{e}_{in}$ and angular frequency $\omega = 2\pi c/\lambda = k_0 c$. For concreteness, the image can be thought as being generated by a plane wave with polarization $\mathbf{e}_{in}$ impinging on an aperture, but the calculations shown here are independent of the way in which the image is created. We define the Fourier transform of the input image as $f_{in}(k_x, k_y) \equiv \int dx dy\, e^{-i(k_x x + k_y y)} E_{in}(x,y)$, where $[k_x, k_y]$ are the in-plane wave vector components. Following standard Fourier optics [2], the image can be decomposed into a bundle of plane waves, each propagating along a direction identified by the polar and azimuthal angles θ and φ. In particular, for an arbitrary image polarization $\mathbf{e}_{in} = [E_{0,x}, E_{0,y}, 0]^T$, the field generated at a point identified by the spherical coordinates $(r, \theta, \phi)$ is given by [3]

$$\mathbf{E}(r,\theta,\phi) = ik_0 \frac{e^{-ik_0 r}}{2\pi r} f_{in}(k_x, k_y)[\mathbf{e}_\theta (E_{0,x}\cos\phi + E_{0,y}\sin\phi) + \mathbf{e}_\phi \cos\theta\, (E_{0,y}\cos\phi - E_{0,x}\sin\phi)] \quad (S1)$$

where we introduced the common transformation of coordinates $[k_x, k_y] = k_0 \sin\theta[\cos\phi, \sin\phi]$. Thus, in the far field of the image ($r \gg \lambda$), the field propagating along each direction $(\theta, \phi)$ can be decomposed (up to an overall constant factor) into s-polarized ($\mathbf{e}_s = \mathbf{e}_\phi$) and p-polarized ($\mathbf{e}_p = \mathbf{e}_\theta$) components,

$$\mathbf{E}_{in}(\theta, \phi) = f_{in}(k_x, k_y)[\mathbf{e}_p E_p(\theta, \phi) + \mathbf{e}_s E_s(\theta, \phi)], \quad (S2)$$

where we defined $E_p(\theta, \phi) \equiv E_{0,x} \cos\phi + E_{0,y} \sin\phi$ and $E_s(\theta, \phi) \equiv \cos\theta (E_{0,y} \cos\phi - E_{0,x} \sin\phi)$. The response of the metasurface can be described by a 2x2 matrix of transfer functions, which dictate the angle-dependent co-polarized and cross-polarized transmission coefficients

$$\overline{\overline{t}}(\theta, \phi) = \begin{pmatrix} t_{ss}(\theta, \phi) & t_{sp}(\theta, \phi) \\ t_{ps}(\theta, \phi) & t_{pp}(\theta, \phi) \end{pmatrix}, \quad (S3)$$

such that the field transmitted through the metasurface at any given angle $(\theta, \phi)$ is

$$\mathbf{E}_{out}(\theta, \phi) = \begin{pmatrix} E_s^{out}(\theta, \phi) \\ E_p^{out}(\theta, \phi) \end{pmatrix} = \overline{\overline{t}}(\theta, \phi) \mathbf{E}_{in}(\theta, \phi) = f_{in}(k_x, k_y) \begin{pmatrix} t_{ss}(\theta, \phi) & t_{sp}(\theta, \phi) \\ t_{ps}(\theta, \phi) & t_{pp}(\theta, \phi) \end{pmatrix} \begin{pmatrix} E_s(\theta, \phi) \\ E_p(\theta, \phi) \end{pmatrix} \quad (S4)$$

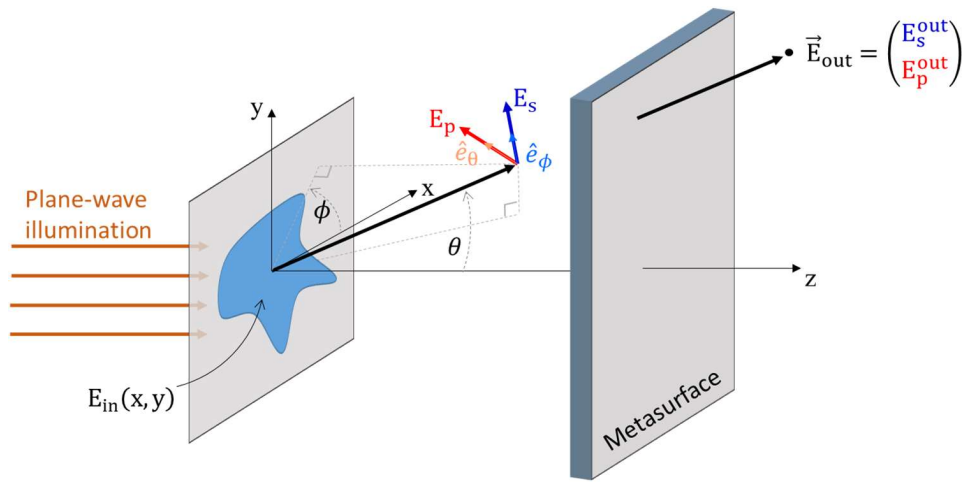

**Figure S3.** Schematic of the fields scattered by the image and filtered by the metasurface. See text for details.

In order to calculate the image generated by this filtered bundle of waves, we project them back onto the z=0 plane. From an experimental point of view, this is equivalent to collecting and re-focusing these waves on a plane placed at a z=4f with a pair of identical lenses with focal length f. Moreover, we transform the field into the x-y polarization basis. This transformation corresponds to the inverse of the plane-wave expansion in Eq. (S1). That is, apart from an overall proportionality factor, the output fields are

$$\begin{pmatrix} E_x^{out}(\theta, \phi) \\ E_y^{out}(\theta, \phi) \end{pmatrix} = \overline{\overline{M}}^{-1}(\theta, \phi) \begin{pmatrix} E_s^{out}(\theta, \phi) \\ E_p^{out}(\theta, \phi) \end{pmatrix} \quad (S5)$$

where $\overline{\overline{M}}^{-1}$ is the inverse of the matrix

$$\overline{\overline{M}}(\theta, \phi) = \begin{pmatrix} \cos\phi & \sin\phi \\ -\cos\theta \sin\phi & \cos\theta \cos\phi \end{pmatrix}. \quad (S6)$$

Finally, the spatially dependent fields $E_x^{out}(x, y)$ and $E_y^{out}(x, y)$, corresponding to the electric field of the filtered image, are obtained via the inverse Fourier transform

$$E_{x/y}^{out}(x, y) = \frac{1}{2\pi} \int dxdy e^{i(k_x x + k_y y)} E_{x/y}^{out}(k_x, k_y), \quad (S7)$$

and the intensity profile is then calculated via $I(x, y) = |E_x^{out}(x, y)|^2 + |E_y^{out}(x, y)|^2$.

In Fig. 1a we considered the case of a polarization-independent edge-detecting metasurface, for which the transferred functions are $t_{ss}(\theta, \phi) = t_{pp}(\theta, \phi) = \sin^2\theta$ and $t_{sp}(\theta, \phi) = t_{ps}(\theta, \phi) = 0$. The input image is a CUNY logo with transversal size equal to about 800 $\lambda$ and pixel size equal to 1.25 $\lambda$. The corresponding edge-detect image (green-colored plot labelled by 'output intensity' in Fig. 1a) was obtained by applying the formulas outlined above, and assuming an x-polarized input image ($\mathbf{e}_{in} = [1, 0, 0]^T$). In Fig. 1a we also showed semi-quantitatively (red-white-blue color-coded image in Fig. 1a) how different parts of the filtered image can be attributed to waves with p- or s-polarization. To create the red-white-blue color-coded image in Fig. 1a, we used the following procedure. We first set the s-polarized transfer function to zero ($t_{ss}(\theta,\phi) = 0$) while maintaining $t_{pp}(\theta,\phi) = \sin^2\theta$. Then, by assuming the same input image and input polarization (x), we calculated the resulting output intensity $I_p(x,y)$. We then considered the opposite case ($t_{ss}(\theta,\phi) = \sin^2\theta$ and $t_{pp}(\theta,\phi) = 0$) and calculated the corresponding output intensity $I_s(x,y)$ with the same input image. Finally, we defined the function $d(x,y) \equiv I_p(x,y) - I_s(x,y)$, which is the quantity plotted in the red-white-blue color-coded image in Fig. 1a. Thus, white colored areas correspond to $d(x,y) = 0$, i.e., to spatial regions of the filtered image that are equally contributed by s- and p-polarized waves. Instead, dark red (dark blue) colors corresponds to spatial regions that are mainly formed by p-polarized (s-polarized) waves.

A similar procedure was used to generate the panels of Figs. 1(b-f). Here, we consider the case of a polarization-dependent metasurface with fixed transfer functions ($t_{ss}(\theta,\phi) = \sin^2\theta$ and $t_{pp}(\theta,\phi) = t_{sp}(\theta,\phi) = t_{ps}(\theta,\phi) = 0$), and we instead vary the input polarization. The input image is shown in Fig. 1b (pixel size = 0.625 $\lambda$). The input polarizations are set to $\mathbf{e}_{in} = [1,0,0]^T$ (Fig. 1c), $\mathbf{e}_{in} = [0,1,0]^T$ (Fig. 1d), $\mathbf{e}_{in} = [1,1,0]^T/\sqrt{2}$ (Fig. 1e) and $\mathbf{e}_{in} = [1,-1,0]^T/\sqrt{2}$ (Fig. 1f), and the corresponding output images are calculated by applying the formulas outlined above. Each of the panels 1(c-f) shows the total intensity $I(x,y) = |E_x^{out}(x,y)|^2 + |E_y^{out}(x,y)|^2$, without any polarization filter.